\documentclass[journal]{IEEEtran}

\usepackage{cite}
\usepackage{amsmath,bm}
\usepackage{graphicx,adjustbox}
\usepackage{subfigure} 
\usepackage{pgf}
\usepackage{gensymb}
\usepackage{amssymb}
\usepackage{url}
\renewcommand\vec{\mathbf}
\DeclareMathOperator*{\argmin}{arg\,min}

\usepackage[scaled]{}
\usepackage{tikz}
\usetikzlibrary{arrows,shapes,positioning, calc,automata}
\newcommand{\MarkRightAngle}[4][.6cm]% #1=size (optional), #2-#4 three points: \angle #2#3#4
{\coordinate (tempa) at ($(#3)!#1!(#2)$);
	\coordinate (tempb) at ($(#3)!#1!(#4)$);
	\coordinate (tempc) at ($(tempa)!0.5!(tempb)$);%midpoint
	\draw (tempa) -- ($(#3)!2!(tempc)$) -- (tempb);
}

\begin{document}

\title{Fast Epipolar Consistency without the Need for Pseudo Matrix Inverses}

\newcommand{\lp}{\vec{l}}				%% L_p	
\newcommand{\Pin}{\vec{P}^+}			%% P+  pseudoinverse
\newcommand{\Pint}{\vec{P}^{+\top}}		%% P+T pseudoinverse transpose
\newcommand{\rax}{\vec{r}}				%% r rot axis homogen
\newcommand{\homo}{\lozenge}			%% homogenize operation
\newcommand{\inhomo}{\blacklozenge}		%% dehomogenization	
\newcommand{\pone}{\mathbb{P}}		%% P
\newcommand{\ptwo}{\mathbb{P}^2}		%% P2
\newcommand{\pthree}{\mathbb{P}^3}		%% P3
\newcommand{\pthreethree}{\mathbb{P}^{3\times3}}		%% P3x3
\newcommand{\pn}{\mathbb{P}^n}		%% Pn
\newcommand{\pnm}{\mathbb{P}^{n-1}}		%% Pn
\newcommand{\rone}{\mathbb{R}}		%% R2
\newcommand{\rtwo}{\mathbb{R}^2}		%% R2
\newcommand{\rthree}{\mathbb{R}^3}	
\newcommand{\rthreethree}{\mathbb{R}^{3\times3}}	%% R3x3
\newcommand{\rthreefour}{\mathbb{R}^{3\times4}}	%% R3x4
\newcommand{\rfour}{\mathbb{R}^4}		%% R4
\newcommand{\rfourfour}{\mathbb{R}^{4\times4}}		%% R4
\newcommand{\rnm}{\mathbb{R}^{n-1}}		%% Rn-1
\newcommand{\rnplusone}{\mathbb{R}^{n+1}}	
\newcommand{\rn}{\mathbb{R}^{n}}		%% Rn-1
\newcommand{\rnmm}{\mathbb{R}^{n-2}}		%% Rn-2
\newcommand{\Pvone}{\vec{P_\text{1}}}	%% P1
\newcommand{\Pvonet}{\vec{P_\text{1}^\top}}	%% P1
\newcommand{\Pvonea}{\vec{P}_\text{1,align}}
\newcommand{\Pvoneaa}{\vec{p}^1_\text{1,align}}
\newcommand{\Pvoneab}{\vec{p}^2_\text{1,align}}
\newcommand{\Pvoneac}{\vec{p}^3_\text{1,align}}	%% P1algn
\newcommand{\Pvtwo}{\vec{P_\text{2}}}
\newcommand{\Pvtwot}{\vec{P_\text{2}^\top}}
\newcommand{\Pvtwoi}{\vec{P_\text{2}^+}}
\newcommand{\Pvtwoit}{\vec{P_\text{2}^{+\top}}}	%% P2
\newcommand{\Pvtwoa}{\vec{P_\text{2,align}}}	%%P2algn
\newcommand{\pu}{\vec{p}_1}				%% p1
\newcommand{\pv}{\vec{p}_2}				%% p2
\newcommand{\pd}{\vec{p}_3}				%% p3
\newcommand{\meet}{\wedge}				%% meet
\newcommand{\join}{\vee}				%% join/incident\vec{\pi}^\infty
\newcommand{\pinfty}{\vec{\pi}^\infty}	%% plane at infity
\newcommand{\radon}{\mathcal{R}} %% Radon transform

\newcommand{\lka}{\vec{l}^\kappa_a}
\newcommand{\lkb}{\vec{l}^\kappa_b}
\newcommand{\lkl}{\vec{l}^\kappa_\lambda}
\newcommand{\Ek}{\vec{E}^\kappa}
\newcommand{\Lk}{\vec{L}^\kappa}
\newcommand{\Ldual}{\vec{L}_K}
\newcommand{\Lkdual}{\vec{L}^\kappa_K}

%\author{\IEEEauthorblockN{Alexander~Preuhs\IEEEauthorrefmark{1}, Michael Manhart\IEEEauthorrefmark{2} and Andreas Maier\IEEEauthorrefmark{1}}\\
%\IEEEauthorblockA{\IEEEauthorrefmark{1}Pattern Recognition Lab, Friedrich-Alexander-Universit\"at Erlangen-N\"urnberg, Erlangen, Germany \\
%\IEEEauthorrefmark{2}Siemens Healthcare GmbH, Forchheim, Germany \\
%Email: alexander.preuhs@fau.de}
%\thanks{A. Preuhs and A. Maier are with the Pattern Recognition Lab, Friedrich-Alexander-Universit\"at Erlangen-N\"urnberg, Erlangen, Germany. M. Manhart is with Siemens Healthcare GmbH, Forchheim, Germany}% <-this % stops a space
%}

\author{Alexander~Preuhs, Michael~Manhart and~Andreas~Maier
		
	\thanks{A. Preuhs and A. Maier are with the Pattern Recognition Lab, Friedrich-Alexander-Universit\"at Erlangen-N\"urnberg, Erlangen, Germany. 	
	
	M. Manhart is with Siemens Healthcare GmbH, Forchheim, Germany. 
	
	Email: alexander.preuhs@fau.de}% <-this % stops a space
}

\maketitle
\pagestyle{empty}
\thispagestyle{empty}
\begin{abstract}
% an abstract should contain:
% What is the Problem (i.e. what are we dealing with) (e.g. motion)
% What is typically done to alleviate this problem (e.g motion compensation)[note this is NOT a lit. review ;)]
% What is this paper about
% What are the results
%\begin{enumerate}
%%	\item what can we do with epipolar consistency
%	\item Grangeats theorem enables comparison of specific line in projections
%	\item Motion compensation and geometric calibration, and in general rigid motion
%	\item optimization of algorithm
%	\item how do we evaluate it -> run time comparison
%\end{enumerate}
Interventional C-arm systems allow flexible \mbox{2-D} imaging of a \mbox{3-D} scene while being capable of cone beam computed tomography. Due to the flexible structure of the \mbox{C-arm}, the rotation speed is limited, increasing the acquisition time compared to conventional computed tomography. Therefore, patient motion frequently occurs during data acquisition inducing inconsistencies in the projection raw data. A framework using Grangeat's theorem and epipolar consistency was successfully applied for compensating rigid motion. This algorithm was efficiently parallelized, however, before each iteration, the pseudo-inverse of each projection matrix must be calculated. We present a geometric modification of the presented algorithm which can be used without a pseudo-inverse. As such, the complete algorithm can be implemented for low-level hardware without the need of a linear algebra package that supports the calculation of matrix inverse. Both algorithms are applied for head motion compensation and the runtime of both is compared. 
\end{abstract}

\IEEEpeerreviewmaketitle

\section{Introduction}
%% an introduction should contain:
%% Typical Problem statement -> why do we have these artifacts (it might be, that it is necessary to further explain why this problem is really a problem ;) e.g. see viewpointplanning)
%% What has been done in Literature to alleviate this problem (this is the literature part)
%% Present the own method, what is does and how it was evaluated

A fundamental assumption in computed tomography (CT) is that the scanned object remains static during the acquisition process. If this assumption cannot be fulfilled, images produced with conventional reconstruction algorithms will suffer from artifacts. Current C-arm CT acquisitions last about 20 seconds. During the acquisition time, involuntary patient motion is often inevitable without patient fixation. However, if the motion can be assumed to be rigid and smooth, a motion compensated reconstruction can be computed by finding the correct geometric correspondence between the motion affected projections and the calibration data. Four categories of compensating motion artifacts have emerged in literature and they can be grouped into approaches using external markers \cite{Kim2013}, image metrics on the reconstruction volume \cite{Sisniega2016}, \mbox{3-D/2-D} registration of the projection data to digitally rendered radiographs from the reconstruction volume \cite{Wein2013,Berger2016} and projection data consistency based metrics \cite{Aichert2015,Debbeler2013,Maass2014,Berger2017}.

In this work, we focus on a consistency method based on the \mbox{3-D} radon transform. The method exploits epipolar geometry to find lines on two detectors corresponding to an epipolar plane. Grangeat's theorem can be used to find a mapping between each epipolar line pair and the \mbox{3-D} radon value corresponding to the epipolar plane \cite{Defrise1994}. This algorithm is denoted as epipolar consistency and was presented by Aichert et al. \cite{Aichert2015}. As the algorithm directly works on the projection domain without the need of a reconstruction, the computational cost is low. It basically consists of comparing corresponding line configurations. This can be accelerated by parallelizing the algorithm using graphics processing units (GPU) \cite{Aichert2016}. 

To apply the algorithm for rigid motion compensation, the consistency between all possible line pairs is evaluated in an iterative optimization process in order to find the set of parameters describing the motion within the scan \cite{Frysch2015,Unberath2017}. In \cite{Aichert2016} before each iteration the pseudo-inverse of the projection matrices must be calculated on the CPU. We propose a geometric modification that allows to calculate corresponding epipolar lines without the need of a pseudo-inverse. 
\section{Methods}
%%%%%%%%%%%%%%%%%%%%%%%%%%%%%%%%%%%%%%%%%%%%%%%%%%%%%%%%%%%%%%%%%%%%%%%%%%%%%%%%%%%%%%%%%%
%%%%%%%%%%%%%%%%%%%%%%%%%%%%%%%%%%%%%%%%%%%%%%%%%%%%%%%%%%%%%%%%%%%%%%%%%%%%%%%%%%%%%%%%%%
%%%%%%%%%%%%%%%
%%%%%%%%%%%%%%%
%%%%%%%%%%%%%%%		Grangeat's Theorem
%%%%%%%%%%%%%%%
%%%%%%%%%%%%%%%%%%%%%%%%%%%%%%%%%%%%%%%%%%%%%%%%%%%%%%%%%%%%%%%%%%%%%%%%%%%%%%%%%%%%%%%%%%

\subsection{Grangeat's Theorem}
In cone-beam CT an X-ray source radially emits photons, that~---~after attenuation~---~are measured at a detector. The attenuation process for a ray can be described by an integral. However, due to the radial structure of the rays, integrating along a detector line does not result in a plane integral of the underlying object $f$, instead it differs by a radial weighting. 

\begin{figure}
	\begin{center}
		\begin{tikzpicture}[scale=0.6]
		\input{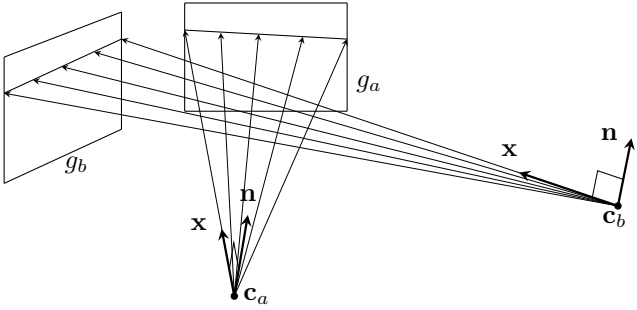}
		\end{tikzpicture}
		\caption{Schematic drawing of a scene including two projections. The vector $\vec{n}$ describes the normal of a radon plane. Several realizations of the vectors $\vec{x}$ are drawn that are perpendicular to $\vec{n}$. The pixel intensity measured at the detector along ray $\vec{x}$ is described by $g_a(\vec{x})$ or $g_b(\vec{x})$, respectively.}
		\label{fig:epipolarGeometry}
	\end{center}
\end{figure}

Grangeat's theorem describes the connection between this weighted integral and a plane integral~---~i.e. the \mbox{3-D} radon value $\radon f(\vec{n},d)$ describing the integral along a plane with normal $\vec{n} \in \mathcal{S}^2$ at distance $d$. Using a derivative operation the radial weighting can be canceled out.
Grangeat defined an intermediate function $S_\lambda(\vec{n})$ that is calculated from the projection data which can be related to the derivative of the \mbox{3-D} radon transform
\begin{equation}
	S_\lambda(\vec{n})=
	\int_{\mathcal{S}^2} \delta'(\vec{x}^\top \vec{n})  g_\lambda(\vec{x})  d \vec{x}=
	\frac{\partial}{\partial d}  \radon f(\vec{n},d)\lvert_{d=\vec{c}^\top_\lambda\vec{n}}
	\enspace,
	\label{eq:intermediate}
\end{equation}
where $g_\lambda(\vec{x})$ describes a single value on the detector with $\lambda$ describing the projection index, $\vec{c}_\lambda$ the source position and $\vec{x}$ a vector from the source to a detector pixel. The geometry for two projections $\lambda=a$ and $\lambda=b$ is visualized in Fig. \ref{fig:epipolarGeometry}. Here $\delta'(\cdot)$ describes the derivative of the Dirac delta distribution%, where the argument $(\vec{x}^\top\vec{n})$ is zero, if the vectors are perpendicular
. A detailed evaluation of Eq. \eqref{eq:intermediate} can be found in \cite{Defrise1994}, and some simplifications are discussed in \cite{Aichert2015}. 

%%%%%
%%%%%	DONE
%%%%%
%\begin{enumerate}
%	\item Grangeat's theorem
%	\item mapping between a line on the detector to radon plane
%\end{enumerate}

%%%%%%%%%%%%%%%%%%%%%%%%%%%%%%%%%%%%%%%%%%%%%%%%%%%%%%%%%%%%%%%%%%%%%%%%%%%%%%%%%%%%%%%%%%
%%%%%%%%%%%%%%%%%%%%%%%%%%%%%%%%%%%%%%%%%%%%%%%%%%%%%%%%%%%%%%%%%%%%%%%%%%%%%%%%%%%%%%%%%%
%%%%%%%%%%%%%%%
%%%%%%%%%%%%%%%
%%%%%%%%%%%%%%%		Epipolar Geometry
%%%%%%%%%%%%%%%
%%%%%%%%%%%%%%%%%%%%%%%%%%%%%%%%%%%%%%%%%%%%%%%%%%%%%%%%%%%%%%%%%%%%%%%%%%%%%%%%%%%%%%%%%%

\subsection{Epipolar Consistency}
\label{sec:epipolar}
It directly follows from Eq. \eqref{eq:intermediate} that two projections $a, b$ must satisfy
\begin{align}
S_a(\vec{n}) =
S_b(\vec{n})
%\int_{\mathcal{S}^2} \delta'(\vec{x}^\top \vec{n})  g_b(\vec{x}) d \vec{x}=
%\int_{\mathcal{S}^2} \delta'(\vec{x}^\top \vec{n})  g_a(\vec{x}) d \vec{x} 
&& \forall \vec{n} \in \mathcal{S}^2: \vec{c}_b^\top \vec{n} = \vec{c}_a^\top \vec{n} 
\enspace.
\label{eq:consistency}
\end{align}
If the geometry information is wrong, e.g. due to rigid object motion, then Eq. \eqref{eq:consistency} will not hold. Thus, we can use it as a measure of inconsistency.
Below we summarize the framework proposed by Aichert et al. \cite{Aichert2015,Aichert2016}, which is used to evaluate the consistency of two views.  

The intermediate function $S_\lambda(\vec{n})$ can be precomputed for each projection. Then, the global indexing by the plane normal $\vec{n}$ can be replaced by a local projection-pair-dependent indexing using a line $\lkl$ defined on the detector described by $g_\lambda$. By epipolar geometry two epipolar lines $\lka$ and $\lkb$ are found that belong to the same epipolar plane $\Ek$~---~i.e. the radon plane. The algorithm starts with a configuration of two projections described by their projection matrix $\vec{P}_a$ and $\vec{P}_b$, respectively. Using these two projection matrices a mapping matrix is  derived that maps an angle $\kappa$ to an epipolar plane $\Ek$. Using the pseudo-inverse the respective epipolar lines $\lka$ and $\lkb$ are computed. The respective values are then used to look up the values at the precomputed intermediate function $S_a$ and $S_b$. This allows the indexing of Eq. \eqref{eq:intermediate} using an angle $\kappa$ and two projection matrices
\begin{align}
	S_a(\kappa, \vec{P}_a, \vec{P}_b) = S_a(\vec{n}) 
	&& \forall \vec{n} \in \mathcal{S}^2: \vec{c}_b^\top \vec{n} = \vec{c}_a^\top \vec{n} 
	\enspace.
\end{align}
To evaluate the consistency of a whole scan, many different views must be compared to each other, while in each two-view comparison a multitude of line pairs are evaluated. As the operations are independent from each other, this can be evaluated in parallel allowing the efficient parallelization of the algorithm using GPUs.  

%%%%%%%%%%%%%%%%%%%%%%%%%%%%%%%%%%%%%%%%%%%%%%%%%%%%%%%%%%%%%%%%%%%%%%%%%%%%%%%%%%%%%%%%%%
%%%%%%%%%%%%%%%%%%%%%%%%%%%%%%%%%%%%%%%%%%%%%%%%%%%%%%%%%%%%%%%%%%%%%%%%%%%%%%%%%%%%%%%%%%
%%%%%%%%%%%%%%%
%%%%%%%%%%%%%%%
%%%%%%%%%%%%%%%		Projective Geometry
%%%%%%%%%%%%%%%
%%%%%%%%%%%%%%%%%%%%%%%%%%%%%%%%%%%%%%%%%%%%%%%%%%%%%%%%%%%%%%%%%%%%%%%%%%%%%%%%%%%%%%%%%%
\subsection{Projective Geometry}
\label{sec:projectiveGeometry}
Projective geometry can be seen as an extension to the common Euclidean geometry. In the context of image reconstructions, projective geometry is mostly used to describe the projection of a world point to a detector. Therefore, a projection matrix is created that performs a projective transformation on a world point. In this context, the world point must be converted to homogeneous coordinates first. 

Homogeneous coordinates are the representation of $n$-dimensional points in the projective space and are written as $(n+1)$-component vectors. In $\pthree$ a point is described by $(x,y,z,w)^\top$, and we can obtain the euclidean representation by dividing with the last component $(x/w, y/w, z/w)^\top$. Similarly, a plane is described by $(a,b,c,d)^\top$. The vector can be understood as the parameters of a Hessian normal form, where the first three components describe the normal of the plane, and $d$ is the scaled distance to the origin. If $a^2+b^2+c^2 = 1$ then $d$ is exactly the signed distance to the origin. The concept that a four-component vector can either be interpreted as a point or a plane is called duality, where we refer to the point interpretation as primal form and the plane interpretation as dual form. 

A special case is the representation of a line in $\pthree$. There is no direct description but we can construct the line as the connection of two points or the intersection of two planes. An intuitive derivation can be found in \cite{Blinn1977}, we only state the relevant result of this derivation. The creation of a line as the incident of two planes $\vec{a}, \vec{b} \in \pthree$ is obtained by
\begin{equation}
\label{eq:meet_pl_pl}
\text{meet}(\vec{a},\vec{b}) =
\vec{L} = 
\begin{pmatrix}
p \\ q \\ r\\ s\\t \\u 
\end{pmatrix} 
= 
\begin{pmatrix}
a_z b_w - a_w b_z\\ a_y b_w - a_w b_y \\ a_y b_z - a_z b_y\\ a_x b_w - a_w b_x\\ a_x b_z - a_z b_x \\ a_x b_y - a_y b_x 
\end{pmatrix} \enspace,
\end{equation}
where the six components of $\vec{L}$ are often referred to as Pl\"ucker coordinates. We can build an anti-symmetric matrix $\Ldual$ from the Pl\"ucker coordinates that represents a line as the intersection of two planes~---~i.e. the dual representation of a line. A point $\vec{x}$ common to a plane $\vec{p}$ and the line $\vec{L}$ can be found by right-multiplication of $\vec{p}$ to $\Ldual$ 
\begin{equation}
\vec{x} = \text{meet}(\vec{L},\vec{p}) =  \Ldual \, \vec{p} =
\begin{pmatrix}
0 & -p & - q & r\\p & 0 & s & -t\\ q & -s & 0& u\\ -r & t & -u & 0
\end{pmatrix}  \, \vec{p} \enspace.
\label{eq:meetlineplane}
\end{equation}
Note that there is also a primal representation of $\vec{L}$ which will not be discussed in this paper.

An extension in the projective geometry is the concept of geometric primitives at infinity. They are regular objects and thus can be handled as any other objects. A point at infinity is defined by a homogeneous coordinate $w=0$. In $\pthree$ the plane at infinity is defined by $\bm{\pi}_\infty = (0,0,0,1)^\top$. All previously introduced equations are also valid for objects at infinity. We could for example use Eq. \eqref{eq:meetlineplane} to find the incident of a line $\vec{L}$ with $\bm{\pi}_\infty$, which will be a point at infinity, where the first three component of that point are the direction of the line. 

An advantage of using projective geometry is the representation of transformations based on matrix multiplication. A point $\vec{x}^\prime$ which is the transformation of the point $\vec{x}$ under $\vec{T}$ is simply found by
\begin{equation}
	\vec{x}^\prime \, =
	 \, \vec{T} \, \vec{x} \enspace.
	 \label{eq:pointtrans}
\end{equation}
The transformation rule for planes can be derived from the property that the distance from a point $\vec{x}$ incident to a plane $\vec{p}$ is zero. The distance between the transformed point $\vec{x}^\prime$ and the plane $\vec{p}^\prime$ will remain zero if they have been transformed under the same transformation $\vec{T}$. It therefore holds that 
\begin{equation}
	\vec{p}^{\prime\top} \vec{x}^\prime
	= 
	\vec{p}^\top \, \vec{x} 
	=
	0
	\enspace .
	\label{eq:zerodist}
\end{equation} 
Solving Eq. \eqref{eq:pointtrans} for $\vec{x}$ and plugging that in Eq. \eqref{eq:zerodist} gives
\begin{equation}	
	 	\vec{p}^{\prime\top} \vec{x}^\prime \,
	 	= 
	 	\vec{p}^\top \,  \vec{T}^{-1} \, \vec{x}^\prime 
	 	= 
	 	\left((\vec{T}^{-1})^\top \, \vec{p} \right)^\top \vec{x}^\prime \enspace ,
\end{equation} 
it directly follows  that  
\begin{equation}
\vec{p}^{\prime} = (\vec{T}^{-1})^\top \, \vec{p}
\label{eq:planetrans}
\end{equation} 
which describes the transformation of planes. The point $\vec{x}$ incident to a plane $\vec{p}$ and a line $\vec{L}$ can be found by right-multiplying the plane to the dual representation of $\vec{L}$ (cf. Eq. \eqref{eq:meetlineplane}). Further, a transformed point $\vec{x}^\prime$ will be incident to the plane $\vec{p}^\prime$ and line $\vec{L}^{\prime}$ if both are transformed under a transformation $\vec{T}$, thus, 
\begin{align}
	\Ldual \, \vec{p} \, = \vec{x}, && \Ldual^{\prime}  \,\vec{p}^\prime \, = \vec{x}^\prime \enspace.
	\label{eq:planemeetline}
\end{align}
When we solve Eq. \eqref{eq:pointtrans} and \eqref{eq:planetrans} for $\vec{x}$ and $\vec{p}$, respectively, we can plug the result in the left part of Eq. \eqref{eq:planemeetline} which results in
\begin{equation}
	\Ldual \vec{T}^{\top} \, \vec{p}^\prime = \, \vec{T}^{-1} \, \vec{x}^\prime
	\,
	\Leftrightarrow
	\,
	 \vec{T} \, \Ldual \vec{T}^{\top} \, \vec{p}^\prime = \,   \vec{x}^\prime \enspace.
\end{equation}
Comparing the result with the right side of Eq. \eqref{eq:planemeetline} it immediately emerges that the line $\vec{L}^{\prime}$ which is the transformation of $\vec{L}$ under $\vec{T}$ can be calculated by
\begin{equation}
\Ldual^{\prime}  \,
=  
\, \vec{T} \, \Ldual \, \vec{T}^\top
\enspace .
\label{eq:transformation_lines}
\end{equation}

%% DONE
%%\begin{enumerate}
%%	\item Line in projective 3-space as meet of two planes
%%	\item Line at infinty
%%\end{enumerate}
%%%%%%%%%%%%%%%%%%%%%%%%%%%%%%%%%%%%%%%%%%%%%%%%%%%%%%%%%%%%%%%%%%%%%%%%%%%%%%%%%%%%%%%%%%
%%%%%%%%%%%%%%%%%%%%%%%%%%%%%%%%%%%%%%%%%%%%%%%%%%%%%%%%%%%%%%%%%%%%%%%%%%%%%%%%%%%%%%%%%%
%%%%%%%%%%%%%%%
%%%%%%%%%%%%%%%
%%%%%%%%%%%%%%%		Optimized Algorithm
%%%%%%%%%%%%%%%
%%%%%%%%%%%%%%%%%%%%%%%%%%%%%%%%%%%%%%%%%%%%%%%%%%%%%%%%%%%%%%%%%%%%%%%%%%%%%%%%%%%%%%%%%%

\subsection{Optimized Algorithm}
The main purpose of the algorithm presented in Section \ref{sec:epipolar} is to find the mapping between two lines $\lka$ and $\lkb$ that can be used to look up the corresponding precomputed values $S_a$ and $S_b$, respectively. This is achieved by first finding epipolar planes $\Ek$ which are then mapped to the corresponding epipolar lines. The algorithm presented in \cite{Aichert2015} makes use of the pseudo-inverse to compute that mapping. However, the calculation of a pseudo-inverse is not supported on many GPUs, and must therefore be done on the CPU beforehand, whereas the rest of the framework is parallelizable. In addition a linear algebra library must be included to support the calculation of pseudo matrix inverses.

We propose a geometric modification that creates the mapping without the need of a pseudo-inverse. As shown in Section \ref{sec:projectiveGeometry}, the transformation rule depends on the object that is to be transformed. It can be seen from Eq. \eqref{eq:transformation_lines} that lines are transformed using the transformation matrix and its transpose. Thus, transforming the plane to a line while preserving the relevant information will make the pseudo-inverse dispensable.

We can achieve this using the concept of infinity. The projective three-space is covered by the infinity plane $\bm{\pi}_\infty = (0,0,0,1)$. Any plane intersects the infinity plane in a line incident to $\bm{\pi}_\infty$ and the plane itself, i.e. a line at infinity. The orientation of the plane is persevered by the direction of the line. In a last step, we can simply use Eq. \eqref{eq:transformation_lines}  to project the line at infinity, resulting in the desired epipolar lines.

Therefore, we start with the epipolar plane $\Ek$. Using Eq. \eqref{eq:meet_pl_pl} we can compute the line at infinity $\Lk$ as the intersection of the epipolar plane with $\bm{\pi}_\infty$
\begin{equation}
	\Lk = \text{meet}(\Ek, \bm{\pi}_\infty) \enspace.
	\label{eq:stepA}
\end{equation}
Using the representation of the line at infinity now allows us to use the transformation rule as described by Eq. \eqref{eq:transformation_lines} to obtain the epipolar line $\lkl$
\begin{equation}
	[\lkl]_\times \, = \, \vec{P_\lambda} \,  \Lkdual \, \vec{P_\lambda}^T \enspace.
	\label{eq:stepB}
\end{equation}
The parameters of $\lkl$ are available from the $3\times3$ skew matrix $S = [\lkl]_\times$ as $\lkl = (S_{12},S_{20},S_{01})^\top$. As a result Eq. \eqref{eq:stepA} and \eqref{eq:stepB} replace the mapping from epipolar planes to lines presented in \cite{Aichert2015} and, therefore, makes the computation of pseudo-inverses unnecessary. The additional cost is the implementation of Eq. \eqref{eq:meet_pl_pl} on the GPU, however, this can be reused to simplify the calculation of source positions. As the three rows of the projection matrix can be interpreted as planes all passing the source, the incident of two of these planes will create a line. Using matrix multiplication (cf. Eq. \ref{eq:meetlineplane}) the source position is then found by the incident of that line with the third plane.

\subsection{Optimization}
If rigid motion occurs during the scan, the calibrated trajectory does not represent the true geometry of the acquired data. In order to restore the true geometry, a rigid transformation $\vec{T}_\lambda$ for each projection matrix $\vec{P}_\lambda$ must be found. The true geometry is expected to have minimal inconsistency. We therefore define the inconsistency between two projections $a$ and $b$ in dependence of the respective rigid transformations $\vec{T}_a$ and $\vec{T}_b$ by
\begin{multline}
	d(\vec{P}_a\vec{T}_a,\vec{P}_b\vec{T}_b) = \frac{1}{N_\kappa}\sum_{k = 0}^{K}\\  \left[S_a(k\Delta\kappa,\vec{P}_a\vec{T}_a,\vec{P}_b\vec{T}_b) - S_b(k\Delta\kappa,\vec{P}_b\vec{T}_b,\vec{P}_a\vec{T}_a)\right]^2 
	\enspace,
\end{multline}
where $N_\kappa$ is the number of epipolar planes that hit both detectors and $K$ is the total number of sampled epipolar planes. The angular step-size is denoted by $\Delta\kappa$. To be more robust for outliers we use the robust Cauchy norm and define the inconsistency of two views by
\begin{equation}
e_{a,b} =\frac{d(\kappa, \vec{P}_a\vec{T}_a,\vec{P}_b\vec{T}_b)}
{1 + \frac{1}{c} \, d(\kappa, \vec{P}_a\vec{T}_a,\vec{P}_b\vec{T}_b)}
\enspace.
\end{equation}
%From Frytsch...The parameter c controls properties of this metric like smoothness, detectability of inconsistencies or consideration of outliers and has to be chosen empirically.
The parameter c controls the penalty and should be selected according to the intensity of the projection images. We denote the vector of rigid transformations $\vec{T} = [T_1, ... T_N]$, with $N$ being the number of projections of the trajectory. The corrected geometry is denoted by $\hat{\vec{T}}$ and found by solving
\begin{equation}
\hat{\vec{T}} = \argmin_\vec{T} \sum_{a,b = 1}^N e_{a,b} \enspace.
\end{equation}
Since motion is expected to be smooth we model each rigid motion parameter in $\vec{T}$ by an Akima spline \cite{Akima1970}. This also allows the reduction of the search space, as we must not find a transformation for each $\lambda$, but only for the nodes of the spline. The optimum is then found using the open source non-linear optimizer JPOP\footnote{\url{https://www5.cs.fau.de/research/software/java-parallel-optimization-package/}} in CONRAD 
\cite{Maier13}.

% Maybe show error grids
\section{Experiments}
To evaluate the proposed method, we have acquired a 200$\degree$ short scan (496 projections) of a head phantom using a robotic C-arm system (Artis zeego, Siemens Healthcare GmbH, Germany). Thereafter, we simulate rigid motion, which is directly incorporated in the projection matrices. This is done using a rigid motion creator\footnote{\url{https://github.com/alPreuhs/MotionCreator}}.

Epipolar consistency is known to produce mostly horizontal epipolar lines in a majority of the projection pairs within a short scan. Only view pairs that are almost opposed to each other present diverging epipolar lines. Motion that is parallel to the epipolar lines is not detectable by the presented consistency measure. Thus, we only concentrate on motion orthogonal to the epipolar lines in all pairs, which is typically denoted as out-plane motion. Defining the rotation axis of the short scan as the $z$-axis, we only simulate translations in $z$-direction. The simulated motion pattern consisting of 17 spline nodes is shown in Fig. \ref{fig:motionplot}.
%With $\vec{z}$ being the rotation axis, the out-plane motion is defined by the two rotations $\vec{R_x}$, $\vec{R_y}$ and a translation $\vec{t_z}$.. The calibrated  The trajectory is then optimized using the conventional method and the presented method, where the calculation of the pseudo-inverse is omitted.

\section{Results}
\begin{figure}
	\centering
	\resizebox{\linewidth}{!}{
		\input{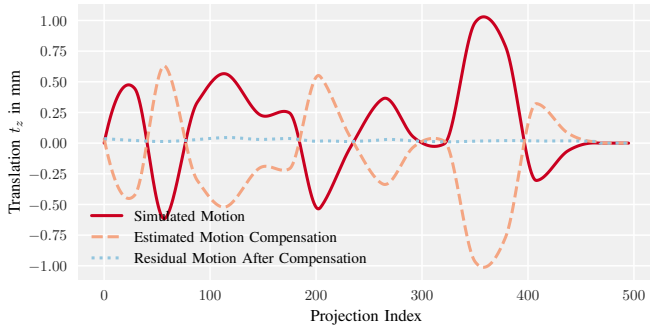}
	}
	\caption{Simulated, estimated and residual motion $t_z$ for each projection.}
	\label{fig:motionplot}
\end{figure}

\begin{figure}
	\centering
	\begin{minipage}{0.3252\linewidth}
	\includegraphics[trim=100 0 40 0,clip,width=\linewidth]{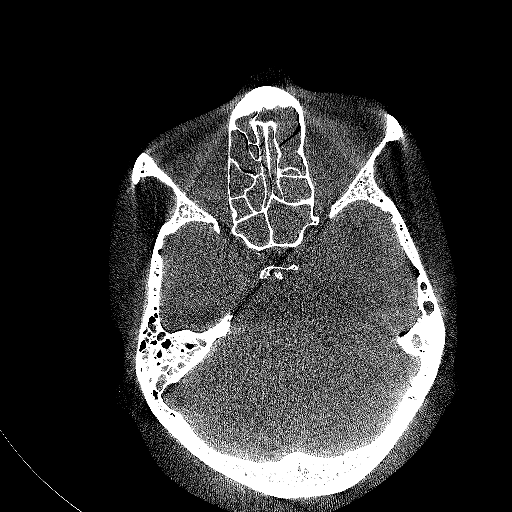}
	\end{minipage}
\begin{minipage}{0.3252\linewidth}
	\includegraphics[trim=100 0 40 0,clip,width=\linewidth]{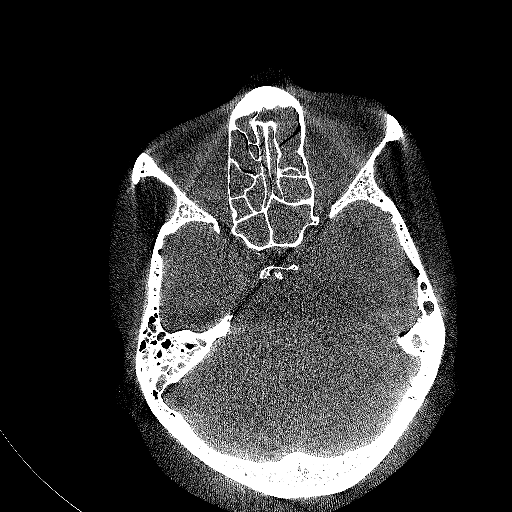}
\end{minipage}
\begin{minipage}{0.3252\linewidth}
	\includegraphics[trim=100 0 40 0,clip,width=\linewidth]{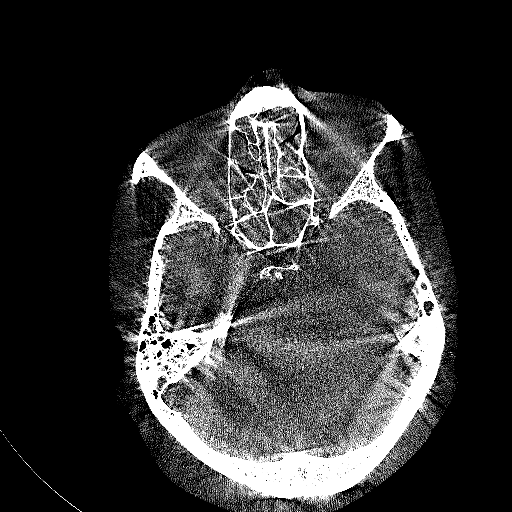}
\end{minipage}
\caption{Central slices of the reconstructed volume HU [-100, 100]. Left: ground truth, mid: with simulated motion after compensation, right: with simulated motion.}
\label{fig:reco}
\end{figure}

The reconstructions of the acquired head phantom is shown in Fig. \ref{fig:reco} for the motion corrupted case (right), the motion compensated case (mid) and the ground truth (left). The corresponding motion is depicted in Fig. \ref{fig:motionplot}. Both algorithms produce the same results, only the runtime is expected to change. By skipping the sequential calculation of pseudo-inverses the runtime could be reduced by $1.29$\% using a standard computer with an Intel Core i7-4910MQ and a NVIDIA Quadro K2100M. The overall runtime for the motion parameter estimation was 841.7 seconds using the proposed modifications and 852.8  seconds if the inverse is pre-calculated before each optimization step. 

%\begin{enumerate}
%	\item Show compensation (which is basically the same)
%	\item Show runtime
%\end{enumerate}
\section{Conclusion and Discussion}
We presented a modification to the algorithm presented in \cite{Aichert2016} which avoids the calculation of inverse projection matrices. This is achieved by transforming the respective epipolar planes to lines at infinity. Lines are transformed~---~in contrast to planes~---~using only the transformation matrix and its transposed. Thus, only the projection matrix and its transposed must be available. 

The runtime could be improved by $1.29$\% using a Java environment. Using more high-level programming languages, e.g. python, the runtime advantage could eventually increase, as more computations can be performed on a dedicated GPU. Furthermore, when implementing the algorithm in low-level programming languages, e.g. C++/CUDA, no linear algebra libraries need to be included. 

%The modification comes with the cost of implementing Eq.   \eqref{eq:meet_pl_pl} and \eqref{eq:meetlineplane} on the GPU

%\begin{enumerate}
%	\item capable of saving time
%	\item integrates to current algorithm
%	\item results will probalby be even more sever when using e.g. python
%\end{enumerate}

%We will decide later
%\subsection*{Disclaimer}
%The concepts and information presented in this paper are based on research %and are not commercially available.

~\\
\textbf{Disclaimer:} The concepts and information presented in this paper are based on research and are not commercially available.

%\newpage



%%Somehow this works now...no clue why and why not before...
\bibliographystyle{IEEEtran}
\bibliography{MyCollection}

\end{document}